# Spin Probe ESR Signature of Freezing in Water: Is it *Global* or *Local*?

Debamalya Banerjee and S. V. Bhat

*Department of Physics, Indian Institute of Science*
*Bangalore 560012, India*

ABSTRACT

First systematic spin probe ESR study of water freezing has been conducted using TEMPOL and TEMPO as the probes. The spin probe signature of the water freezing has been described in terms of the collapse of narrow triplet spectrum into a single broad line. This spin probe signature of freezing has been observed at an anomalously low temperature when a milimoler solution of TEMPOL is slowly cooled from room temperature. A systematic observation has revealed a spin probe concentration dependence of these freezing and respective melting points. These results can be explained in terms of localization of spin probe and liquid water, most probably in the interstices of ice grains, in an ice matrix. The lowering of spin probe freezing point, along with the secondary evidences, like spin probe concentration dependence of peak-to-peak width in frozen limit signal, indicates a possible size dependence of these localizations/entrapments with spin probe concentration. A weak concentration dependence of spin probe assisted freezing and melting points, which has been observed for TEMPO in comparison to TEMPOL, indicates different natures of interactions with water of these two probes. This view is also supported by the relaxation behavior of the two probes.

I. INTRODUCTION

Water is the most common solvent on earth used widely in different types of biological and chemical systems. A considerable abundance of water is commonly encountered in almost all sorts of biological liquid samples and many of these samples are chosen to be studied by the technique of spin probe (or spin label if the sample contains a nytroxyl side chain) ESR. On the other hand, supercooling, vitrification and glass transition of water, pose continuous scientific challenges which could not be studied experimentally by means of conventional thermodynamic measurement techniques like differential scanning calorimetry (DSC) and dielectric relaxation. Recently we have addressed the question of vitrification and glass transition using spin probe ESR[1]. A further analysis of the data shows the existence of a small supercooled water fraction in a polycrystalline ice matrix consisting of both high and low density phases of liquid water (namely HDL and LDL)[2].

Albeit of fundamental interest in many respect, very little of water phase transitions had been studied by spin probe ESR. Here we choose to omit the report related to the glass transition temperature - as glassy phase of matter is not a stable thermodynamic phase, glass transition cannot be considered as a true phase transition - of bulk water by the same method[9]. Except for some rigorous studies on liquid crystals[3] and polycrystalline solids[14], thermodynamic phase transitions are not very well studied with the technique of spin probe ESR.

Electron spin resonance (ESR) is renowned for its high sensitivity and its ability to probe fast dynamics ($\sim 10^{-12}$ sec or less) attributed to the high frequency (GHz) microwave used for the detection. If the system subjected to ESR study does not have a paramagnetic centre, the external tracers, namely the *spin probes* are added to it. Although the use of trapped metallic ions as the spin probe is known for some lipids and liquid crystals[3], the probes are usually organic molecules with a nitroxyl ($-N=O$) group attached to one side. The unpaired electron associated with the oxygen molecule of the nitroxyl radical is responsible for resonance absorption of energy when exposed to adequate magnetic field in the presence of microwave of appropriate frequency. For a long time the spin probe method has been extensively used to study the dynamics of various types of systems. The variety of studies by spin probe includes the systems like solid TMCB matrix[5] and oil-water interface[6] to the glass transition and related dynamics of polymer matrix[7], nematic phase of liquid crystal[8] and even electron magnetic resonance imaging of biological species[9]. Recently we have demonstrated its usefulness in the studies of glass transition in polymer electrolytes[10] and molecular liquid like water[1].

We chose the most common (and probably the best understood) phase transition in water, *the ice-water transition of bulk water*, where the isotropic liquid changes into hexagonal ice $I_h$ crystals and the volume of the system increases. The ice-water equilibrium temperature for bulk water is the well known 273.16 K (= $T_f$). However, almost all crystallization experiments exhibit some amount of supercooling as can be seen from figure 1. A DSC experiment for bulk water with a cooling rate of 3 K/min detects the freezing transition at around 260 K and the onset of melting starts soon after the temperature crosses $T_f$. Though the amount of supercooling monotonically increases with reduction of the size of water body and enhanced cooling rate, bulk water could not be supercooled below 235 K (commonly referred as the homogeneous nucleation temperature $T_h$) under any circumstances[11]. However, the situation is different for water in nanoconfinement. A recent report shows water to remain liquid up to ~190 K when confined to 1.4 nm and 1.8 nm channels of MNC- 41[12]. Moreover, by conducting relaxation measurements on this deeply supercooled water, the existence of fragile-to-strong dynamic crossover (FSC) was also observed. Across this transition water changes from a fragile liquid which has an energy landscape consist of multiple megabasins into a strong liquid with single megabasin[13]. Experimentally determined value of ~225 K is close to the value $T_{FSC}$ ~ 228 K proposed by Ito *et al.* for water[14].

To our knowledge, there are four previous reports of spin probe ESR studies on bulk water freezing. One of these, John S. Leigh, Jr. *et al.* states the freezing cause solute-solvent segregation and these segregated solutes do not freeze unless some eutectic composition is reached[15]. They have studied the freezing of a variety of inorganic and organic paramagnetic species including the probe TEMPO. The report by O. I. Mikhalev *et al.* describes the effect of inorganic salts of various concentrations on this segregation for the spin probe TEMPOL in the frozen aqueous salt solutions[16]. In these reports, the signature of the bulk water freezing in the spectra of TEMPO and TEMPOL were described in terms of the collapse of liquid state narrow triplet spectrum into a single broad line as observed in the figure 2, last panel of this report as well. No specific quantitative value of the temperature at which this change took place was mentioned as the experiments in both of the reports were carried out at the liquid nitrogen temperature of 77 K. Bhat *et al.*, had attributed this change in the spectrum to aggregation of the spin probe molecules thrown out from ice crystals[1]. In the report of Jas Singh *et al.*, N-oxyl-4,4-dimethyloxazolidine 5-ketostearic acid (Formula Weight 398) was chosen as the spin probe. Freezing was induced by the cooling in small steps, and the broadening of the linewidth was observed at -2 $^0$C (271 K) [17].

We chose TEMPO (2,2,6,6-tetramethylpiperidin - 1- oxyl, molecular weight 156.24) and TEMPOL (4-hydroxy-2,2,6,6-tetramethylpiperidin - 1- yloxy, molecular weight 172.25), the most common of the spin probes, to study bulk water freezing. The choice was made based on the solubility of these two probes in water as well as their moderate size ($r_{TEMPO}$ ~ *0.3 nm, $r_{TEMPOL}$ ~ 0.34 nm*, to be compared with $r_{H2O}$ ~ *0.14 nm*). As we report below, the results of ESR of water of these spin probes are counter-intuitive.

## II. EXPERIMENT

The samples were prepared by adding TEMPO (98%, Aldrich) and TEMPOL (97%, Aldrich) into triple distilled, de-ionized water in the concentrations range of $10^{-4} - 10^{-3}$ Mole. A few µl of the sample was then taken into a capillary of 1mm inner diameter (typical sample mass ~ 8 mg), which was then mounted with the help of a quartz tube in the Oxford continuous flow, helium cooled cryostat attached to the spectrometer. The sample was then slowly cooled down (typical cooling rate 20 K/hr) from room temperature. Temperature was controlled by with a Oxford PID type controller with thermocouple sensor. After reaching the lowest temperature of the experiment, the sample was slowly re-heated back to the room temperature. EPR spectrum at different temperatures was recorded in derivative mode with the help of a Bruker EMX X-band CW EPR spectrometer. Before every temperature run a minimum of 10 minutes of equilibration period was allowed and the spectrometer was re-tuned for the temperature effect on the cavity. A very low power of 20 db (~ 2 mW) and modulation field of 1G was used throughout the experiments to ensure no saturation or line broadening effect on the spectrum.

## III. RESULTS AND DISCUSSIONS

Before the ESR experiments, ~8.1 x $10^{-4}$ M aqueous solution (the highest spin probe concentration used for the current set of experiments) of the probes were subjected to Differential Scanning Calorimetry (DSC) with a cooling/heating ramp of 3 K/min. No significant shifts of freezing or melting signature for these spin probe solutions were observed from that of the corresponding pure water signatures (Figure 1). But, when a TEMPOL solution of same concentration was cooled slowly, the EPR signal remained unaffected across the 273 K landmark and even across the 235 K limit of the bulk water supercooling. The spin probe freezing signature was finally observed between 207K and 205K (Figure 2a). On re-heating, the narrow triplet spectrum reappeared around 245 K. For a TEMPO solution of similar concentration, the corresponding temperatures were recorded as 242K and 265K respectively. One should note the presence of the narrow wings in the frozen limit TEMPO signal at 242 K which almost disappeared at 230 K (Figure 2b, last panel). On reheating, these wings grew stronger and merged with the re-apparent narrow triplet signal above ~265 K. Similar observation was also made in few other slow cooling experiments for both the probes. On repetitions of the experiment, these spin probe assisted freezing and respective melting temperatures turned out to be consistently reproducible. We then investigated a few lower concentrations (lowest one being 1.45x $10^{-4}$ M) for both the probes by means of spin probe ESR and DSC as well. For all the samples significant lowering of these spin probe assisted freezing and respective melting points were observed from the corresponding DSC values. On the other hand, these DSC values show negligible spin probe concentration dependence (data not shown). When plotted as a function of concentration, a systematic trend of these spin probe assisted 'freezing' and respective 'melting' temperatures towards the corresponding DSC values for bulk water in the limit of zero spin probe concentration was revealed for both TEMPO and TEMPOL (Figure 3). The other behaviors of interest were the temperature dependence of the spin probe re-orientation correlation time and the double integrated (absolute) intensity across the transition. The former was estimated from the measured lineshape parameters using the approximate formula

$$\tau_c = [\{(h_{+1}/h_{-1})^{0.5} -1\} \Delta H_{+1}]/1.2 \times 10^{10} \text{ s} \quad \ldots\ldots\ldots\ldots\ldots\ldots\ldots \quad (1)$$

from Freed and Frankel[18], where $h_{+1}$, $h_{-1}$ and $\Delta H_{+1}$ are the heights of the first and third line of the narrow triplet signal and the width of the first line in gauss respectively. The correlation times for TEMPOL, when plotted against inverse temperature (Arrhenius plot) (Figure 4a), showed no significant deviations across the important temperatures like 273 K ($T_f$), 235 K ($T_h$) and 228K ($T_{FSC}$). For TEMPO, the typical correlation time behavior showed a transition knee in the vicinity of 260K (Figure 4b), to be compared with the DSC freezing point. The latter, namely the double integrated or absolute intensity showed a sudden, discontinuous jump for both the probes across the temperature where the lineshape changes from narrow triplet to a broad one (Figure 5). This sudden change in intensity was attributed to the dynamical arrest of the spin probe molecules across the transition. The net magnetization - proportional to the absolute intensity- of the system is the ensemble average over the spins of individual probe particles. The reduced dynamics and enhanced dipolar coupling has strong effects on this ensemble average, thus changing the intensity value abruptly across the transition. Moreover, the *peak-to-peak* width of the broad, frozen limit "crystalline phase" signals for TEMPOL are also seen to follow systematic concentration dependence (Figure 5a, inset). The average (of several repetition of the same experiment) values of this width ranges between 20G – 14G to be compared with 11.2G of dry, bulk TEMPOL (data not shown). A similar observation for TEMPO could not be conducted because of the low signal to noise ration of the frozen limit signals.

Established theory for freezing point depression (colligative properties) cannot account for the huge change in spin probe assisted 'freezing' and 'melting' points. For example, Roult's Law[19] of freezing point depression which signifies lowering of freezing point, estimates a $\Delta T$ of ~ 0.015 K for the highest spin probe concentration used which helps us little to understand the results. Moreover, the other observations related to the strong concentration dependence of spin probe transition temperatures and the presence of occasional 'wings' in the frozen limit broad signal could not be explained in terms of the eutectic composition of the segregated solute as proposed in reference 21. Though the idea of eutectic composition should lead to a single eutectic freezing temperature for all the starting concentrations, in the present study it was not observed. So, reasonable explanations are required with the help of a better model. On the other hand, the concentration dependence of SPESR freezing and melting points are steeper in case of TEMPOL as compared to TEMPO. This indicates that the nature of interaction with water is different for these two probes, a fact also reflected in different nature of the relaxation pattern.

As discussed earlier, recent studies on water confined, in 1.4 nm and 1.8 nm nanochannels, show that on cooling water retains its liquid form down to 190 K[12]. So, a reasonable explanation of the low spin probe freezing point can be given in terms of the confinement effect and local behavior of the probe. Moreover, the variation of the *peak-to-peak* width of the TEMPOL signal in the frozen sample with concentration, and their corresponding larger widths in comparison to respective signal widths of dry bulk TEMPOL also indicate that the probe molecules are confined in smaller sizes. Similar ESR line broadening for solid phase TEMPOL was previously observed with the probe confined in SWCN. The broadening was attributed to the reduced exchange interaction between probe molecules due to the confinement in a low dimensional system[20]. On the basis of the earlier reports and present observations, we have tried to explain the phenomena with the aid of following model.

When an assembly of spin probe molecules dissolved in water is cooled slowly, the probe molecules below the water freezing point get trapped - along with liquid water molecules - inside tiny islands located most probably in the interstices between the ice grains.

This has also been proposed in literature a number of times[15, 16, 21], but detailed discussions about the entrapments were lacking. To our understanding, the average size of these islands is a function of the solute (spin probe in the present case) concentration and these islands have negligible variation of size for a given concentration. This entrapment has impressive effects on the freezing signature of spin probes. Due to confinement effect of small sizes, the trapped liquid can stay in liquid phase well below $T_f$. Consequently, the probe molecules can continue there dynamic re-orientation to response with narrow triplet spectrum in EPR measurement. Finally, when the crystallization of this confined liquid takes place, the probe molecules segregate together (driven by the hydrogen bond network of water which forbids impurities from interfering with the ice structure) resulting huge line broadening due to dipolar coupling and frozen dynamics. Subsequently, the triplet spectrum changes into one broad line. The average size of this entrapment shell and the number of entrapped spin probe molecules is most likely a decreasing function of spin probe concentration. On the other hand, the total number of such islands is likely to increase with concentration. One might explain the higher depression in spin probe freezing point with the enhanced concentration, as described in figure 3, in terms of this smaller size of the entrapment shell (Figure 6). Similarly, the higher *peak-to-peak* width of the frozen limit signal for higher concentration (Figure 5a, inset) may also be explained successfully as for higher probe concentration, entrapments in reduced numbers of particle may lead to a reduced exchange narrowing in frozen state, thus to the broadening of the signal solid phase signal.

As discussed earlier, in the report of Singh *et al.*[17] the spin probe signature of freezing appears at a temperature very close to the crystallization temperature. This might be due to the large size and molecular mass of the probe used. Due to large formula weight (398) the chance of this probe molecule to interfere with the hydrogen bond network of water and being thrown out easily of the host matrix is high.

The frozen limit crystallization phase signals of TEMPO with 'wings', as observed in figure 1b, were analyzed with the help of rigorous computer simulation. The details of simulation may be found elsewhere[22]. The overall lineshapes were obtained with the help of weighted superposition of a signal broadened due to frozen dynamics and dipolar interaction to a triplet spectrum of dilute limit, mobile free radical executing isotropic rotation. Figure 7 shows both the experimental and simulated signals for 242 K (a) and 230 K (b) respectively. The isotropic hyperfine coupling constant of the narrow signal for both the temperatures were determined to be 15.66 G, comparable with the corresponding liquid regime value of 17.3 G. The changing relative weight of the broad to narrow signal ($w_b/w_n$) with temperature, from

~89 at 242 K to ~194 at 230K, indicates a decreasing intensity of the narrow triplet signal signifying a considerable lowering of the liquid fraction in the sample. This also can be explained in terms of the above mentioned confinement model if we assume the probability of formation of a few confinements in a relatively smaller size than the average for a given concentration (Figure 8). Smaller size of these entrapment shells leads to a lower SPESR freezing temperature that $T_f^{ESR}$; $T_f^{ESR}$ being the temperature where all other domains of sizes close to the average size freezes. The freezing of these smaller domains takes place at a temperature lower than $T_f^{ESR}$ as we have seen the narrow wings to almost disappear at 230K. On reheating, these smaller entrapments gather mobility at a temperature lower than the temperature where majority of the average sized domains becomes mobile, *i.e.* the SPESR melting point $T_m^{ESR}$ of the system. This is apparent from the observation that these wings become more intense with increasing temperature and eventually merge with the re-apparent, narrow triplet. We might as well consider the probability of confinements in larger than average size islands. On cooling, these domains should freeze at a temperature higher than $T_f^{ESR}$. But because of the low intensity of the solid phase broad signal in comparison with the liquid state triplet, the freezing signature could never be distinguished from the background.

The proposed model is valid at least for the spin probe concentration range of interest ($10^{-3} – 10^{-4}$ M). In the limit of zero spin probe concentration the dilution of probe will prevent cluster formation of any kind and the system will eventually follow the phase diagram of pure water. The higher concentration limit, on the other hand, will lead to a thermodynamically detectable freezing point depression and the system will follow the thermodynamics of a two phase aqueous system. The upper and lower limit of spin probe concentration for the validity of this model is a subject matter of further investigation. However, the importances of our understanding lie in the following facts: firstly, the concentration range of interest is widely used for almost all types of spin probe study. And secondly, milimolar aqueous solutions and emulsions are not uncommon in chemical and biological studies. This model may be used to explain the localized nature of those systems at temperatures below the thermodynamic freezing points.

Encouraged by the extent of freezing point depression, one might estimate these confinement domains to be of nano scale. But with a caveat, that for the same concentrations TEMPO shows a much lower depression of freezing point comparable to that of micron sized droplets of water. But it can be safely stated that these domains occupy only a fraction of the entire sample small enough to remain undetected by a thermodynamic measurement technique such as DSC. Although, the important question remains is at what temperature does this global to local decoupling *i.e.*, the spin probe entrapments takes place. Unlike TEMPOL, the presence of an inflection point in the relaxation pattern of TEMPO encourages us to think that for TEMPO it takes place at a temperature close to the DSC freezing point. But the experiments with TEMPOL leave no hint. The difference in these two probes is also evident in the relaxation pattern. The –OH group present in the TEMPOL has higher ability to form hydrogen bond. So, a possibility of TEMPOL getting more involved into the hydrogen bond network of water than TEMPO seems reasonable. To support this view the solubility of the two probes can be compared as TEMPOL is more soluble in water than TEMPO. The exact model to explain the differences in freezing/melting point depression and relaxation pattern for these two probes is still lacking. A reasonable way of investigation might be to compare a few other standard first order phase transitions (like solid-nematic-isotropic phase transition of some standard liquid crystals) with the help of these two probes.

In summary, a systematic investigation on supercooling of water along with spin probe solution by the method of ESR reveals the presence of entrapped liquid water, most probably at the interstices between the ice grains, in a crystalline ice matrix. These entrapments have their changing nature with the change of spin probe molecules (TEMPO and TEMPOL in the present case) and there varied concentrations. It has been observed that the two probe molecules interact differently with water. This is reflected in the concentration dependence of the spin probe assisted freezing and melting points and their relaxation behavior.

FIGURE CAPTIONS:

Figure 1: DSC signature of crystallization and melting for pure water (panel 1) and highest concentrations of TEMPOL and TEMPO used for the present set of experiments (panel 2 and 3 respectively). No significant deviation of DSC freezing and melting signature is observed for the spin probe solutions from the corresponding pure water values. All experiments were performed with a heating/cooling ramp of 3 K/min. The dotted vertical line marks the 273 K ($0^o$ C) value.

Figure 2: Temperature variation EPR spectrum for 0.81 mM solution of (a) TEMPOL, (b) TEMPO. The spin probe signature of liquid-crystalline phase transition took place at ~205 K for TEMPOL and ~242K for TEMPO. Note the narrow 'wings' present at the crystalline phase signals of the last two plots, left panel of (b). The sharp signal at ~3340G in frozen limit signals of (a) is from the glass capillary used in the experiment.

Figure 3: (Colour online) Spin probe concentration dependence of ESR freezing and melting temperatures for (a) TEMPOL and (b) TEMPO. The star and diamond indicates the DSC freezing and melting points for pure water respectively. The solid lines are guide to the eye.

Figure 4: Arrhenius representation of typical spin probe rotational correlation time data on cooling for (a) TEMPOL and (b) TEMPO. The TEMPOL data show no significant variation within experimental error in the temperature range. Note the transition knee around 260K for TEMPO.

Figure 5: (Colour online) Typical double integrated intensity across the ESR freezing transition for (a) TEMPOL and (b) TEMPO. The spin probe ESR spectra of liquid and crystalline regime are incorporated for convenience. Inset in (a): The peak-to-peak width of the frozen limit signal for TEMPOL. Enhanced width indicates reduces exchange narrowing of aggregated spin probe molecules. Solid lines are guide to the eye.

Figure 6: The model describing the presence of liquid fraction in bulk water below its thermodynamic freezing point. The entrapped liquid/spin probes located within solid ice matrix are responsible for the narrow triplet spectrum well below the freezing point of bulk water. Also describes the change in the confinement pattern with changing spin probe concentration. Whereas relatively low concentration leads to a larger confinement shells with more spin probe particle trapped inside (a), the higher concentration results in more numbers of confinement shells of reduced size and less number of probes trapped in it (b). Due to smaller sizes, the entrapments in the higher concentration solution freeze at a temperature $T_1^{ESR}$ lower than the corresponding temperature of $T_2^{ESR}$ for the lower concentration one.

Figure 7: (Colour online) Two ESR spectra corresponding to 242 K (a) and 230 K (b) from figure 1(b) (for the TEMPO solution) are compared with corresponding simulations. The increment of relative weight of components with decreasing temperature indicates a reduction of narrow component.

Figure 8: (Colour online) The 'wings' in the crystalline phase signal may be explained in terms of confinements in smaller than average size. The smaller entrapments remain in liquid form below the temperature where all the average-seized entrapments freeze. The smaller, still-liquid entrapments are responsible for narrow wings as indicated.

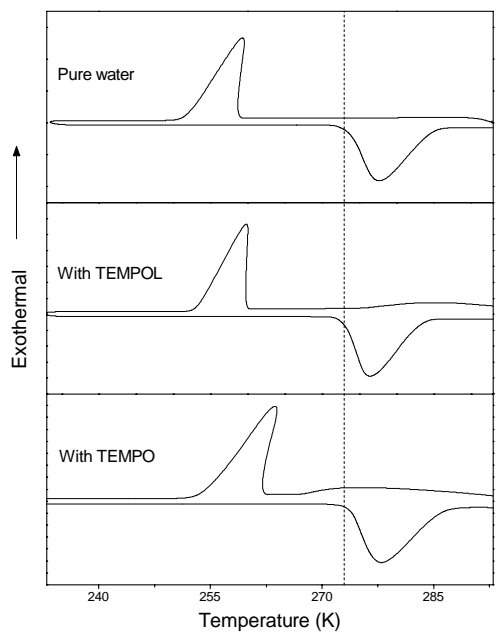

Figure 1

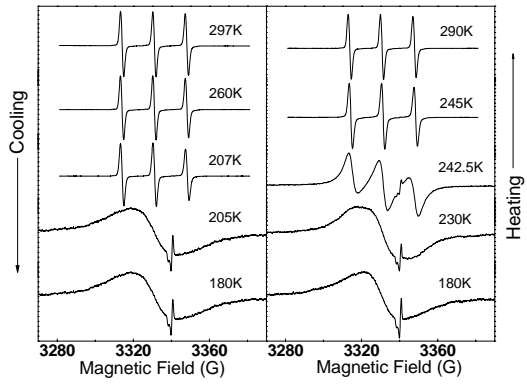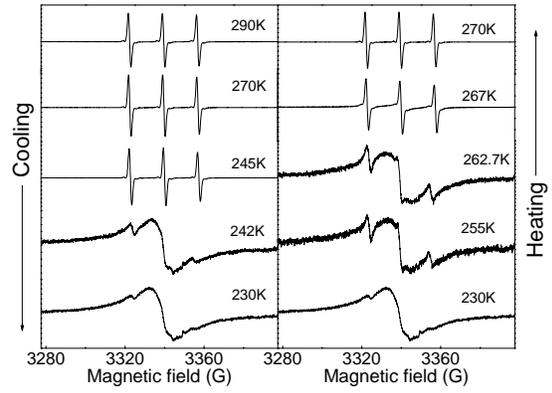

Figure 2

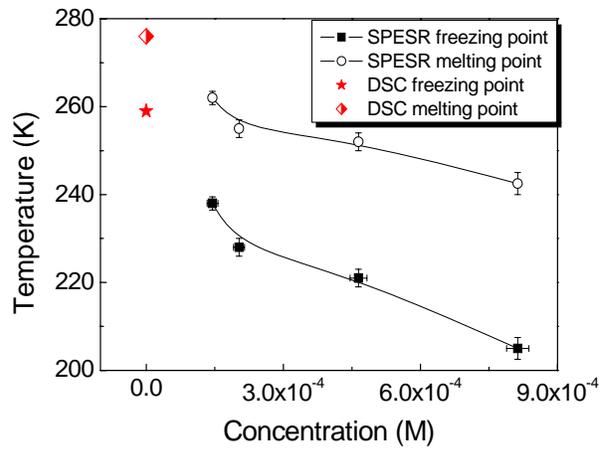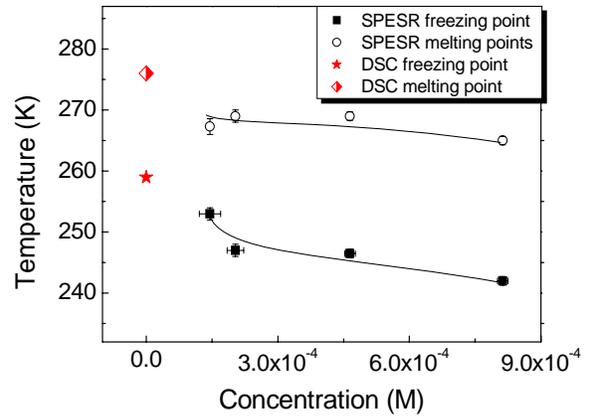

(a)  (b)

Figure 3

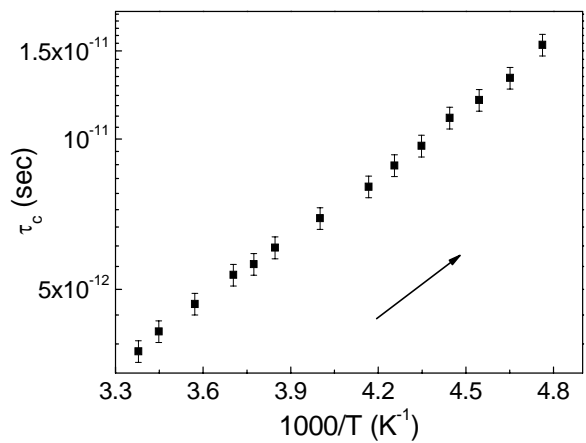 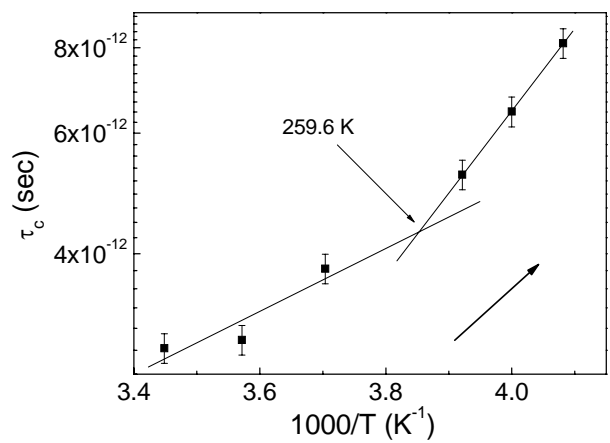

(a) (b)

Figure 4

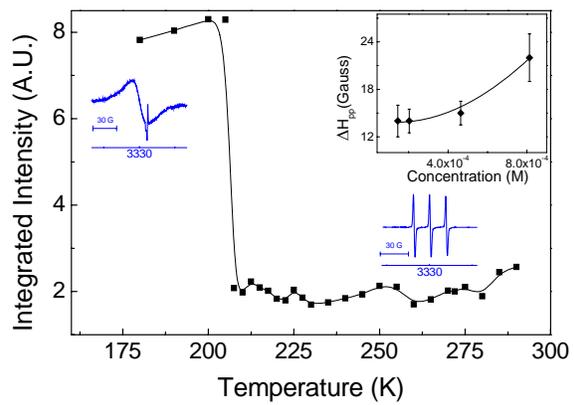 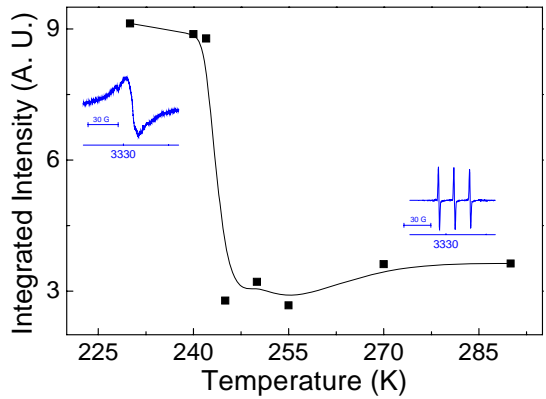

(a)             (b)

Figure 5

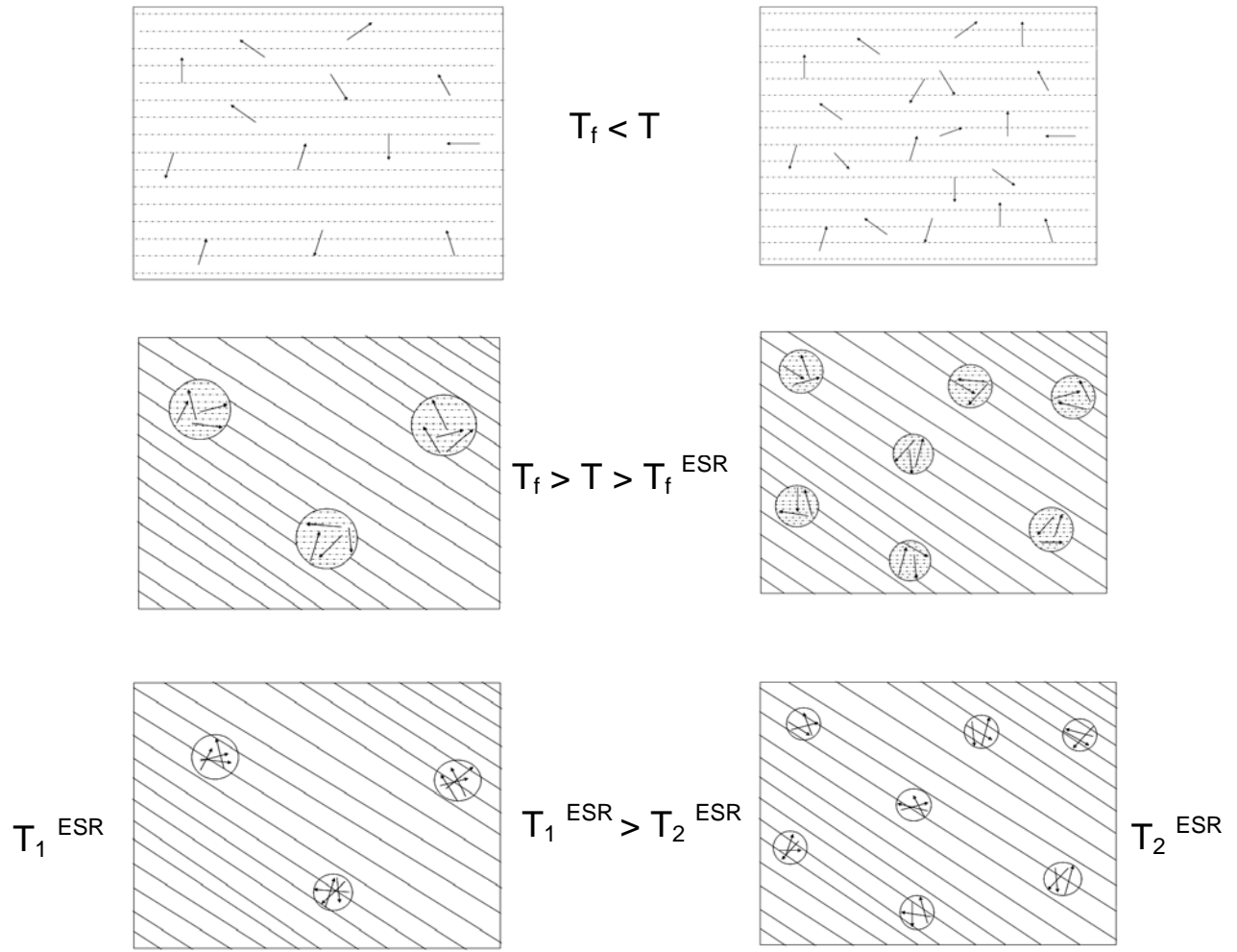

Figure 6

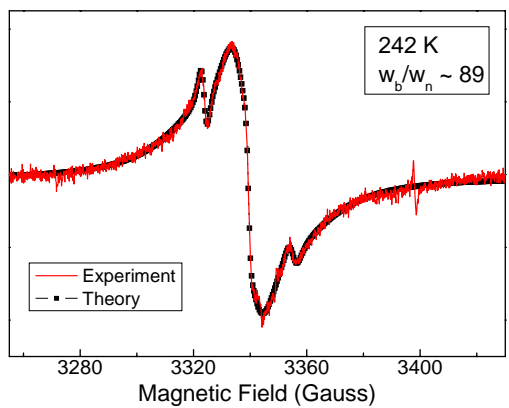 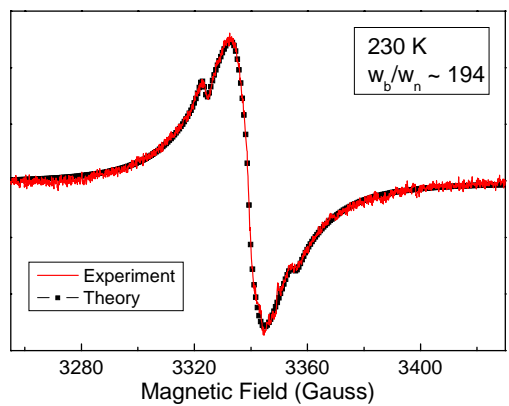

Figure 7

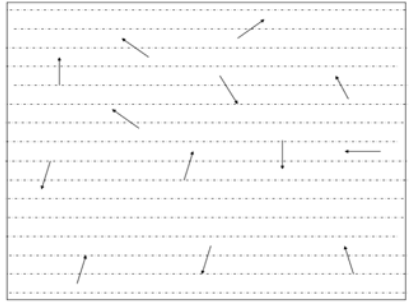 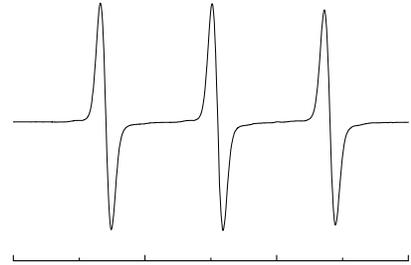

$T > T_f$

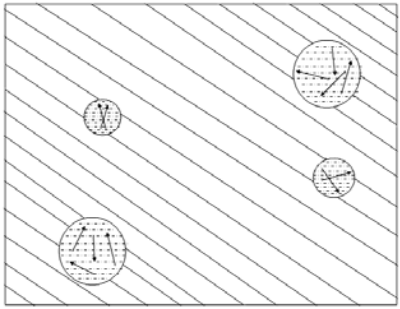 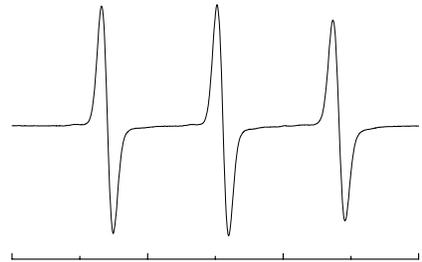

$T_f > T > T_f^{ESR}$

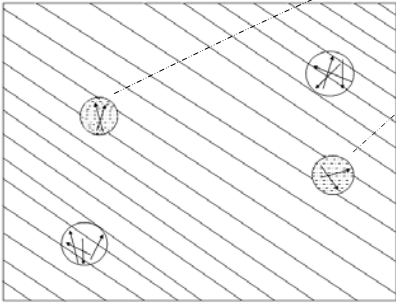 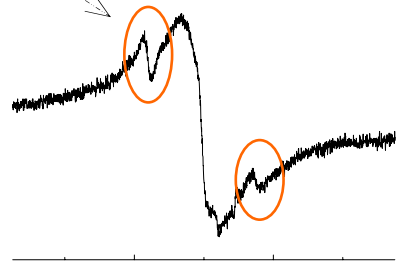

$T < T_f^{ESR}$

Figure 8